\documentclass{article}
\usepackage{spconf,amsmath,graphicx,hyperref,booktabs}
\usepackage{amsmath}
\usepackage{amssymb}
\usepackage{multirow}

\title{MeanVC: Lightweight and Streaming Zero-Shot Voice Conversion via Mean Flows}

\name{
Guobin Ma$^{1}$
Jixun Yao$^{1}$
Ziqian Ning$^{1}$
Yuepeng Jiang$^{1}$ 
Lingxin Xiong$^{2}$
Lei Xie$^{1}\sthanks{Corresponding author.}$
Pengcheng Zhu$^{2}\footnotemark[1]$
}
\address{
$^{1}$ Audio, Speech and Language Processing Group (ASLP@NPU), School of Computer Science, \\
Northwestern Polytechnical University, Xi’an, China \\
$^{2}$ Geely Automobile Research Institute (Ningbo) Company Ltd, Ningbo, China  \\
}

\begin{document}
\ninept
\maketitle
\begin{abstract}
Zero-shot voice conversion (VC) aims to transfer timbre from a source speaker to any unseen target speaker while preserving linguistic content. 
Growing application scenarios demand models with streaming inference capabilities. 
This has created a pressing need for models that are simultaneously fast, lightweight, and high-fidelity.
However, existing streaming methods typically rely on either autoregressive (AR) or non-autoregressive (NAR) frameworks, which either require large parameter sizes to achieve strong performance or struggle to generalize to unseen speakers. 
In this study, we propose MeanVC, a lightweight and streaming zero-shot VC approach.
MeanVC introduces a diffusion transformer with a chunk-wise autoregressive denoising strategy, combining the strengths of both AR and NAR paradigms for efficient streaming processing.
By introducing mean flows, MeanVC regresses the average velocity field during training, enabling zero-shot VC with superior speech quality and speaker similarity in a single sampling step by directly mapping from the start to the endpoint of the flow trajectory.
Additionally, we incorporate diffusion adversarial post-training to mitigate over-smoothing and further enhance speech quality.
Experimental results demonstrate that MeanVC significantly outperforms existing zero-shot streaming VC systems, achieving superior conversion quality with higher efficiency and significantly fewer parameters.
Audio demos and code are publicly available at https://aslp-lab.github.io/MeanVC.
\end{abstract}
\begin{keywords}
streaming voice conversion, zero-shot, mean flows, adversarial post-training
\end{keywords}

\vspace{-4pt}
\section{Introduction}
\label{sec:intro}
\vspace{-2pt}

Zero-shot voice conversion (VC) aims to transfer the timbre from the source speaker to any unseen speaker while keeping the linguistic content unchanged.
This capability has profound implications for a wide array of applications, including movie dubbing~\cite{bgmvc, expressive-vc}, privacy protection~\cite{voiceprivacy2022}, personalized pronunciation correction, and the creation of expressive virtual avatars.
Benefiting from modern deep learning based approaches, recent voice conversion models~\cite{DBLP:conf/icassp/HussainNHLG23, DBLP:conf/icassp/LiLL23, DBLP:conf/icassp/SEF-VC, DBLP:conf/icassp/LuoD24,DBLP:conf/icassp/KimKCNMC25,DBLP:conf/icassp/ChoiP25} have achieved remarkable performance in terms of naturalness and speaker similarity. 
However, as the demand for real-time processing in various application scenarios increases, the computational cost of high-performing models becomes a significant bottleneck for practical deployment. 
This has created a pressing need for streaming models that are simultaneously fast (low-latency), lightweight (low computational overhead), and high-fidelity (high audio quality and speaker similarity).

Recent streaming zero-shot VC studies have explored two main architectural paradigms: autoregressive (AR) and non-autoregressive (NAR) frameworks. A representative AR-based streaming VC system is StreamVoice~\cite{DBLP:conf/acl/WangCW0W24}, which employs a fully causal, context-aware language model. It alternates between semantic bottleneck features and acoustic codec tokens at each AR step, supported by an auxiliary acoustic predictor, to enable streaming zero-shot VC without future look-ahead. Building on this design, StreamVoice+~\cite{DBLP:journals/spl/WangCWXW24} introduces a learnable semantic encoder and a connector, while adapting the backbone via LoRA~\cite{DBLP:conf/iclr/HuSWALWWC22}. It further improves quality through a residual-bottleneck connector, forming an end-to-end streaming pipeline. Despite their high-fidelity output, these models fundamentally compromise on the other two critical requirements. Their sequential decoding process inherently leads to high latency, failing the 'fast' criterion. Moreover, their massive model sizes (101M for StreamVoice and 153M for StreamVoice+) make them far from 'lightweight,' resulting in prohibitive computational costs that render real-time inference on standard CPUs impractical.

To mitigate the inherent latency of AR models, NAR frameworks have been explored for their potential in parallel generation. DualVC2~\cite{DBLP:conf/icassp/NingJ0WY0B24}, for instance, presents a lightweight NAR model based on the conformer architecture, which accelerates inference by incorporating dynamic chunk training. While it achieves faster speeds, its reliance on conventional architectures  limits its generalization capability to unseen speakers in zero-shot scenarios. Another NAR approach adapts powerful diffusion models for zero-shot VC. Seed-VC~\cite{DBLP:journals/corr/seedvc} employs a diffusion transformer (DiT)~\cite{DBLP:conf/iccv/PeeblesX23} architecture within a sliding-window pipeline. Although this mechanism enables streaming VC, the fixed-size windows can disrupt long-term contextual dependencies and speaker characteristics. 
Thus, while NAR frameworks successfully address the 'fast' requirement, they often do so by sacrificing 'high-fidelity,' either through poor generalization in simpler models or context fragmentation in more complex ones.
This leaves a clear gap in the field: a framework that can unite the speed of NAR models with the quality and consistency of their AR counterparts.

\begin{figure*}[!ht]
  \centering
  \includegraphics{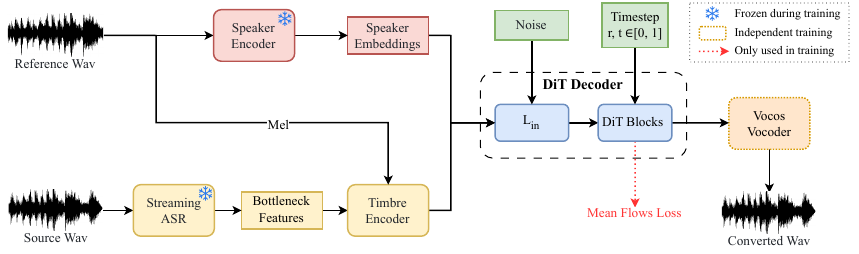}
  \caption{Overall architecture of our proposed MeanVC.}
  \label{fig:overview}
  \vspace{-12pt}
\end{figure*}

To address these challenges, we propose MeanVC, a lightweight framework for streaming zero-shot VC that leverages the advantages of both AR and NAR paradigms while addressing their respective limitations. 
Specifically, MeanVC employs a chunk-wise autoregressive denoising strategy, which processes audio in individual chunks while explicitly conditioning each chunk on the previous one, thereby preserving long-term speaker consistency. 
To overcome the latency of iterative diffusion samplers, we introduce mean flows~\cite{DBLP:journals/corr/MeanFlows}, a powerful method that enables the diffusion model to generate high-quality results with 1-NFE (number of function evaluations).  We integrate mean flows into the DiT-based decoder. By training the model to regress the average velocity field of the probability flow, MeanVC can synthesize a high-quality spectrogram using 1-NFE.
Finally, we incorporate a diffusion adversarial post-training (DAPT) stage to mitigate over-smoothing and further enhance speech quality.
With merely \textbf{14M} parameters, MeanVC achieves superior audio quality and efficiency. Both subjective and objective evaluations demonstrate that MeanVC significantly outperforms existing streaming VC systems. 
Audio demos and code are publicly available at https://aslp-lab.github.io/MeanVC.

\vspace{-4pt}
\section{MEANVC}
\label{sec:methods}
\vspace{-2pt}

\vspace{-3pt}
\subsection{Overview}

As illustrated in Fig.~\ref{fig:overview}, MeanVC is built on the widely used recognition-synthesis framework~\cite{DBLP:journals/taslp/ZhangLD20}, consisting of a streaming automatic speech recognition (ASR) module, a speaker encoder, a timbre encoder, a DiT decoder, and a vocoder. First, the pre-trained streaming ASR model extracts bottleneck features (BNFs) from the source waveform. 
Subsequently, these BNFs are fed into the timbre encoder, which fuses them with fine-grained timbre information extracted from a reference mel-spectrogram to produce timbre BNFs. 
Meanwhile, speaker embeddings are extracted from a reference waveform via a pre-trained speaker encoder. The DiT decoder generates the converted mel-spectrogram conditioned on both the speaker embeddings and the timbre BNFs. During generation, the source mel-spectrogram is concatenated at the front to serve as a cache, enabling chunk-wise autoregressive denoising and producing the converted spectrogram with the target speaker’s timbre.

\vspace{-3pt}
\subsection{Chunk-wise Autoregressive Denoising}
\label{ssec:Chunk-wise Autoregressive Denoising}

To enable streaming conversion, we adopt a chunk-wise autoregressive approach that divides the BNFs into smaller chunks and applies a chunk-wise causal mask inspired by MoonCast~\cite{DBLP:journals/corr/MoonCast}. This mask strategy allows each chunk to access the historical context from previously generated chunks, ensuring consistency across chunks. Furthermore, we build on conditional flow matching (CFM) to learn optimal transport paths between the two distributions.
Given data $x \sim p_{\text{data}}(x)$, prior $\epsilon \sim p_{\text{prior}}(\epsilon)$, the optimal transport flow path can be constructed as: $z_{t}=(1-t)x + t\epsilon$, and the conditional velocity is thus given by $v_t=dz_{t}/dt=\epsilon - x$. During training, the objective is to learn a neural network $f_{\theta}$ that minimizes the following CFM loss: $\mathcal{L}_{\text{CFM}}(\theta)=\mathbb{E}_{t, x, \epsilon}\left\| f_{\theta}\left(t, z_{t}\right)-v_{t}\right\| ^{2}$.

Our DiT decoder is conditioned on timbre BNFs and speaker embeddings to generate the mel-spectrogram from random Gaussian noise. Specifically, we repeat the speaker embeddings to match the length of the timbre BNFs and concatenate them to form the conditioning inputs for the model. We select chunk $i$ to be generated, along with all preceding chunks, where $i$ is an integer index and $i \in [0, N]$ with $N$ being the total number of chunks. Let $M_i$ denote the mel-spectrogram of chunk $i$ and $C_i$ denote its corresponding conditioning input. The mel-spectrograms of the preceding chunks are collectively denoted as $M_{<i}$. 
The flow-matching approach involves a forward process to add noise to the data and a backward process to remove the noise in reverse. During the training phase, the forward process takes as input the noisy chunk $Z_i(t) = (1 - t)M_i + t\xi$, where the clean mel-spectrogram $M_i$ is mixed with Gaussian noise $\xi \sim \mathcal{N}(0, 1)$ at timestep $t \in [0, 1]$. The flow-matching model $f_\theta$ parameterized by $\theta$, learns the velocity field mapping $f_\theta(t, Z_i(t)) = dZ_i(t)/{dt} = \xi - M_i$ conditioned on $C_i$.

\begin{figure}[h]
  \centering
  \hspace{-15pt}
  \includegraphics[width=0.35\textwidth]{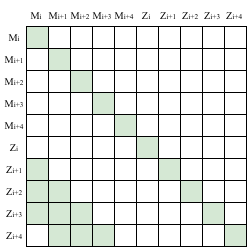}
  \caption{Chunk-wise causal mask in the DiT decoder. In this example, each noisy mel-spectrogram chunk $Z_i$ can attend to up to 3 preceding clean mel-spectrogram chunks $M_j$ (where $j \in [i-3, i-1]$) and itself. The green cells indicate allowed attention, while the white cells indicate restricted attention.
}
  \label{fig:chunk_ar_attn}
  \vspace{-10pt}
\end{figure}

The mel-spectrograms of preceding chunks, denoted as $M_{<i}$, are used as prompts for in-context learning. We apply a chunk-wise causal mask, as illustrated in Fig.~\ref{fig:chunk_ar_attn}. We denote the clean mel-spectrogram chunk as $M_i$ and the noisy mel-spectrogram chunk used for generation as $Z_i$. During training, all clean chunks $M_i$ are concatenated before all noisy chunks $Z_i$ to form a complete sequence of $2N$ chunks. 
The chunk-wise attention mask is defined as follows: 1) Full attention is allowed within each chunk. 2) For clean chunks where $i \in [0, N)$, each chunk $M_i$ can only attend to itself. 3) For noisy chunks where $i \in [N, 2N)$, each chunk $Z_i$ can attend to preceding clean chunks $M_j$ where $j \in [\max(0, i-K), i-1]$ as well as to itself, where $K$ represents the maximum number of preceding clean chunks accessible for conditioning. We find that when using extremely short chunks (e.g., 160 ms), allowing access to too many historical chunks causes the model to over-rely on historical context rather than the current chunk $Z_i$, ultimately degrading performance.

\vspace{-3pt}
\subsection{Mean Flows for Generative Modeling}
\label{ssec:Mean Flows for Generative Modeling}

CFM typically requires multiple sampling steps to solve the ordinary differential equation (ODE), and the number of function evaluations directly impacts inference efficiency. With just 1-NFE, the quality of generated spectrograms degrades catastrophically. To address this limitation, we adopt the approach proposed in mean flows~\cite{DBLP:journals/corr/MeanFlows}, which regresses the average velocity field during training to enable high-quality synthesis with only 1-NFE. 
Specifically, given a time interval $[r, t]$, the displacement on the ODE trajectory within it is $\int_r^t v(z_\tau, \tau) d\tau$. The average velocity is then defined as: $u(z_t, r, t) \triangleq \frac{1}{t-r} \int_{r}^{t} v(z_\tau, \tau) d\tau$. By differentiating both sides with respect to $t$ and subsequently rearranging the resulting terms, the relationship between $u$ and $t$ can be derived, yielding the mean flows identity:

\begin{equation}
    u(z_t, r, t) = v(z_t, t) - (t - r) \frac{d}{dt} u(z_t, r, t).
\end{equation}
By utilizing the Jacobian-vector product (JVP), the total derivative can be expressed as: $\frac{d}{dt} u(z_t, r, t) = v(z_t, t) \partial_z u + \partial_t u $. The velocity $v(z_t, t)$ is the marginal velocity in flow matching. We follow \cite{DBLP:conf/iclr/LipmanCBNL23} to replace it with the conditional velocity $v_t=\epsilon - x$. With this, the target field for training is defined as $u_{\text{tgt}} = v_t - (t - r)( v_t \partial_z u_\theta + \partial_t u_\theta )$. We then encourage $f_\theta$ to satisfy this field by minimizing the mean flows objective:
\begin{equation}
    \mathcal{L}_{\text{MF}}(\theta) = \mathbb{E}_{t, r, x, \epsilon} \left\| f_\theta(z_t, r, t) - \text{sg}(u_{\text{tgt}}) \right\|^2,
\end{equation}
where $\text{sg}(\cdot)$ denotes the stop-gradient operation. Note, when $t = r$, the mean flows objective reduces to the standard flow matching objective.

During sampling, the time integral in CFM can be substituted with the average velocity, resulting in:
\begin{equation}
    z_r = z_t - (t - r) u(z_t, r, t).
\end{equation}
For 1-NFE sampling, we have $z_0 = z_1 - f_\theta(z_1, 0, 1)$, where $z_1 = \epsilon \sim p_{\text{prior}}(\epsilon)$.

\vspace{-3pt}
\subsection{Diffusion Adversarial Post-Training}
\label{ssec:Diffusion Adversarial Post-Training}

Mel-spectrograms generated by flow matching models often suffer from over-smoothing artifacts. While conventional multi-length discriminators~\cite{DBLP:journals/corr/HiFiSinger} are employed, we find that high-frequency components in generated mel-spectrograms still exhibit artifacts such as bright lines.
Inspired by \cite{DBLP:journals/corr/APT}, we leverage the generator architecture as the discriminator backbone to address this issue. Specifically, we employ the DiT model as generator $G$, which generates spectrograms with 1-NFE. The DiT weights are used to initialize the backbone of the discriminator $D$. After training, the discriminator outputs a logit that distinguishes between real samples $x$ and generated samples $\hat{x}$. 
To achieve this, we modify the DiT architecture by introducing cross-attention-only Transformer blocks at the second and fourth layers of the 4-layer DiT backbone. These blocks use a learnable vector as the query and perform cross-attention with the backbone's latent representations as keys and values. The outputs are concatenated channel-wise to form a global feature vector, which is then projected to produce a scalar logit output.

The training objectives of DAPT are defined as follows:
\begin{equation}
    \mathcal{L}_{\text{Adv}}(G) = \mathbb{E}_{c, \epsilon} \left\| D(G(\epsilon, c), c) - 1 \right\|^2,
\end{equation}
\begin{equation}
    \mathcal{L}_{\text{Adv}}(D) = \mathbb{E}_{c, x} \left\| D(x, c) - 1 \right\|^2 +  \mathbb{E}_{c, \epsilon} \left\| D(G(\epsilon, c), c) \right\|^2,
\end{equation}
where $c$ denotes the conditioning input (timbre BNFs and speaker embeddings). The generator $G$ learns to produce realistic samples through 1-NFE generation, while the discriminator $D$ learns to distinguish between real and generated samples.

\vspace{-6pt}
\section{EXPERIMENTS}
\label{sec:EXPERIMENT}
\vspace{-2pt}

\subsection{Experiment Setup}
\label{ssec:Experiment Setup}

\noindent\textbf{Dataset.}
We train MeanVC on the open-source Emilia~\cite{DBLP:conf/slt/Emilia} corpus. Specifically, we use DNSMOS~\cite{DBLP:conf/icassp/Dnsmos} to filter out audio with DNSMOS scores below 3.4, retaining 10,000 hours of mandarin data. All audio files are resampled to 16 kHz for VC training.
For zero-shot evaluation, we use the Mandarin subset of the Seed-TTS test set~\cite{DBLP:journals/corr/Seed-tts}, comprising 2,018 source–target testing pairs. For known speakers evaluation, we fine-tune MeanVC on Aishell3~\cite{DBLP:conf/interspeech/AISHELL-3}, selecting four speakers as targets for evaluation. We use 100 clean and 100 noisy clips as source recordings.
To extract semantic features, we incorporate the streaming ASR model Fast-U2++~\cite{DBLP:conf/icassp/Fast-U2++}, implemented by WeNet~\cite{DBLP:conf/interspeech/WeNet} and trained on WenetSpeech~\cite{DBLP:conf/icassp/WENETSPEECH}.

\noindent\textbf{Implementation Details.}
MeanVC consists of 14M parameters. The DiT decoder employs four DiT blocks, each with a hidden size of 512 and two attention heads. The timbre encoder incorporates two cross-attention modules, with a hidden size of 256 and four attention heads.
Fast-U2++ utilizes a 160\,ms chunk size for streaming inference and compresses 16\,kHz waveforms into semantic features with a 40\,ms frame length.
Speaker embeddings are extracted using ECAPA-TDNN~\cite{DBLP:conf/interspeech/ECAPA-TDNN}.
We employ Vocos~\cite{DBLP:conf/iclr/Vocos} as the vocoder to convert mel-spectrograms to high-fidelity 16\,kHz speech waveforms.

\begin{table*}[ht]
  \centering
  \caption{Zero-shot performance (unseen speakers). The RTF is calculated only for the VC module. The Latency represents the delay of a single chunk (160 ms) within the entire pipeline. The best-performing metrics are indicated in bold, while the second-best are underlined.}
  \label{tab:zeroshot}
  \small
  \renewcommand\arraystretch{1.22}
  \resizebox{0.95\textwidth}{!}{
    \begin{tabular}{ l c c c c c c c c c c}
    \toprule
    \multirow{2}{*}{Method}
      & \multicolumn{3}{c}{Quality} 
      & & \multicolumn{2}{c}{Similarity}
      & & \multicolumn{3}{c}{Efficiency}
      \\ \cline{2-4} \cline{6-7} \cline{9-11}
        & NMOS$\uparrow$ & DNSMOS$\uparrow$ & CER(\%)$\downarrow$ & & SMOS$\uparrow$ & SSIM$\uparrow$ & & Parameters(M)  & RTF$\downarrow$   & Latency(ms)$\downarrow$ \\ 
      \midrule
      GT         & 4.04$\pm$0.05 & 3.79  & 1.36     &   &  -    & -   &   & -  & - & -     \\
      \midrule
      StreamVoice         & \underline{3.81$\pm$0.06} & 3.67  & 9.32     &   &  \underline{3.67$\pm$0.05}    & 0.543 &  & 101  & 13.632    & 2379.52  \\
      
      Seed-VC              & 3.76$\pm$0.07 & \textbf{3.84}  & \underline{6.03}     &   &  3.62$\pm$0.09    & \underline{0.582}  &  & 25   & \underline{7.039}  & \underline{1990.72}     \\
      
      MeanVC              & \textbf{3.82$\pm$0.05} & \underline{3.76}  & \textbf{5.01}    &   &  \textbf{3.87$\pm$0.06}    & \textbf{0.687}  &  & 14   & \textbf{0.136}   & \textbf{211.52} \\
      \bottomrule
    \end{tabular}
    }
    \vspace{-13pt}
\end{table*}

\noindent\textbf{Evaluation Metrics.}
For subjective evaluation, we employ naturalness mean opinion score (NMOS) and speaker similarity mean opinion score (SMOS) as subjective metrics, calculated with 95\% confidence intervals. We randomly select 100 testing pairs and recruit 15 listeners to conduct the assessments. 
For objective evaluation, we measure speech intelligibility using character error rate (CER) computed by an ASR model\footnote{https://huggingface.co/funasr/paraformer-zh}. Speaker similarity (SSIM) is computed using a WavLM-finetuned speaker verification model\footnote{https://github.com/BytedanceSpeech/seed-tts-eval} to measure the similarity between converted and target speaker speech. Additionally, we employ DNSMOS\footnote{https://github.com/microsoft/DNS-Challenge/tree/master/DNSMOS} to assess speech quality.
For streaming performance evaluation, we use real-time factor (RTF) and latency metrics. RTF represents the ratio of model inference time to input speech duration, indicating computational efficiency. We evaluate both RTF and latency of all VC models on a single-core AMD EPYC 7542 CPU with single-threaded execution.

\noindent\textbf{Baseline Systems.}
To validate the performance of MeanVC, we compare MeanVC against the following baseline systems:
\textit{StreamVoice}~\cite{DBLP:conf/acl/WangCW0W24}, a language model-based system that employs fully causal context-awareness. It alternately processes semantic and acoustic features to achieve streaming zero-shot voice conversion;
\textit{Seed-VC}~\cite{DBLP:journals/corr/seedvc}, a NAR system based on diffusion models. It employs an external timbre shifter to mitigate timbre leakage and achieves streaming zero-shot voice conversion through a sliding-window pipeline;
\textit{DualVC2}~\cite{DBLP:conf/icassp/NingJ0WY0B24}, a lightweight NAR system based on the conformer architecture. It uses classic non-causal convolutions and dynamic masking convolutions to achieve streaming any-to-many voice conversion.

\vspace{-8pt}
\subsection{Evaluation on Zero-shot VC}
\label{ssec:Zero-shot Evaluation}
\vspace{-2pt}

To evaluate zero-shot VC performance, we benchmark MeanVC against Seed-VC and StreamVoice. As shown in Table~\ref{tab:zeroshot}, MeanVC outperforms all baseline systems on both subjective SMOS and NMOS, and achieves the best performance on objective metrics CER and SSIM, highlighting its effectiveness in zero-shot VC. However, MeanVC lags behind Seed-VC in DNSMOS, which we attribute to its relatively smaller parameter size compared to Seed-VC.

Our VC module consists of only 14M parameters, achieving an RTF of 0.136 on a single-core, single-threaded setup of an AMD EPYC 7542 CPU, significantly lower than baseline systems. To meet real-time requirements, RTF must remain below 1.0. Our complete pipeline includes the ASR encoder with an RTF of 0.120 and the vocoder with an RTF of 0.066, achieving an overall RTF of 0.322 and satisfying real-time constraints. 
With a 160 ms chunk size, the model's inference latency is 51.52 ms. The total pipeline latency, including the chunk duration, is 211.52 ms (160 ms + 51.52 ms), significantly outperforming baseline systems. Notably, all efficiency measurements for the experiments were obtained without optimizations such as quantization.

\begin{table}[ht]
\vspace{-12pt}
  \centering
  \caption{Intra-dataset performance (seen speakers)}
  \label{tab:indataset}
  \fontsize{8}{10}\selectfont 
  \setlength{\tabcolsep}{0.6mm}  
  \renewcommand\arraystretch{1.22}
  \resizebox{0.48\textwidth}{!}{
    \begin{tabular}{ l c c c c c c c}
    \toprule
    \multirow{2}{*}{Method}
      & \multicolumn{3}{c}{Clean} 
      & & \multicolumn{3}{c}{Noise}
      \\ \cline{2-4} \cline{6-8}
        & \scriptsize{DNSMOS}$\uparrow$  & \scriptsize{CER(\%)}$\downarrow$ & \scriptsize{SSIM}$\uparrow$ & & \scriptsize{DNSMOS}$\uparrow$  & \scriptsize{CER(\%)}$\downarrow$ & \scriptsize{SSIM}$\uparrow$ \\ 
      \midrule
      GT                             & 3.64     & 0.37  & - &   & 2.86& 2.92& -\\
      \midrule
      DualVC2                        & 3.63     & 3.84  & 0.659 &    & 3.47& 16.28& 0.562\\
      MeanVC                         & 3.69     & 3.33  & 0.681 &  & 3.56& 12.84& 0.633\\
      $\quad  +$Tuning             & 3.74    & 3.09 & 0.696 &    & 3.61 & 10.81& 0.657\\
      \bottomrule
    \end{tabular}
  }
  \vspace{-10pt}
\end{table}

\vspace{-8pt}
\subsection{Evaluation on Intra-dataset}
\label{ssec:Intra-dataset Evaluation}
\vspace{-1pt}

To gain deeper insights into MeanVC's performance, we explore it on known speakers by fine-tuning it on Aishell3, selecting four speakers as targets for evaluation. We select DualVC2 for comparison, an any-to-many VC model. Specifically, we use a configuration with parameter size comparable to MeanVC and train DualVC2 from scratch on both Emilia (using the same filtered training data as MeanVC) and Aishell3.
As shown in Table~\ref{tab:indataset}, MeanVC outperforms DualVC2 across all metrics even without fine-tuning, with its advantages becoming particularly evident under noisy input conditions. This demonstrates MeanVC's superior robustness and strong conversion capabilities. With available utterances of target speakers, fine-tuning significantly improves performance. This indicates our system can be easily applied to various scenarios with or without the utterances of target speakers.

\vspace{-7pt}
\subsection{Ablation Study}
\label{ssec:Ablation Study}
\vspace{-1pt}

To validate the contribution of MeanVC's key components and analyze the impact of key hyperparameters, we conducted a series of ablation studies, with results detailed in Table~\ref{tab:ablation}.
First, we assessed the model's core modules. Ablating the DAPT stage resulted in a tangible degradation across all metrics, confirming its crucial role in mitigating over-smoothing and enhancing speech quality. Similarly, disabling the clean mel-spectrogram cache led to a significant performance drop, particularly in CER, which underscores the mechanism's importance in reducing the train-inference mismatch to improve speech intelligibility.
Next, we investigated the critical trade-off between latency and performance by varying the streaming chunk size from our 160 ms default. Reducing the chunk size to 80 ms halves the processing latency, offering a faster response. However, this came at the cost of degraded performance, especially in CER and DNSMOS, likely due to insufficient contextual information for robust modeling. Conversely, increasing the chunk size to 200 ms provided more acoustic context and yielded substantial improvements across all metrics, but at the expense of a 25\% increase in overall latency. These findings explicitly demonstrate the performance-latency trade-off inherent in the streaming pipeline and validate our choice of 160 ms as a well-balanced configuration that achieves strong performance without incurring excessive delay.

\begin{table}[ht]
\vspace{-12pt}
  \centering
  \caption{Results of ablation studies. The baseline MeanVC uses a chunk size of 160ms.}
  \label{tab:ablation}
  \fontsize{8}{10}\selectfont 
  \renewcommand\arraystretch{1.12}
  \resizebox{0.48\textwidth}{!}{
    \begin{tabular}{ l c c c}
    \toprule
    Method    & DNSMOS$\uparrow$ & CER(\%)$\downarrow$  & SSIM$\uparrow$ \\ 
      \midrule
      MeanVC        & 3.76 & 5.01  & 0.687       \\
      \midrule  
      \quad w/o DAPT                    & 3.68  & 5.86  & 0.673       \\
      \quad w/o clean chunks           & 3.71  & 5.97  & 0.677       \\
      \quad w/ chunk size (80ms)             & 3.56  & 9.97  & 0.619       \\
      \quad w/ chunk size (200ms)            & 3.83  & 4.42  & 0.700       \\
      \bottomrule
    \end{tabular}
  }
  \vspace{-12pt}
\end{table}

\vspace{-4pt}
\section{CONCLUSIONS}
\label{sec:CONCLUSIONS}
\vspace{-2pt}

In this study, we introduce MeanVC, a lightweight and efficient framework for streaming zero-shot voice conversion.
By integrating the advantages of both autoregressive and non-autoregressive approaches, MeanVC employs a chunk-wise autoregressive denoising strategy, leverages mean flows for efficient single-step spectrogram synthesis, and incorporates diffusion adversarial post-training to enhance speech naturalness.
Despite its compact architecture of only 14M parameters, MeanVC achieves superior performance in both audio quality and computational efficiency. Our extensive evaluations demonstrate that MeanVC outperforms existing streaming VC systems.

\bibliographystyle{IEEEbib}
\bibliography{strings,refs}

@inproceedings{DBLP:conf/acl/WangCW0W24,
  author       = {Zhichao Wang and
                  Yuanzhe Chen and
                  Xinsheng Wang and
                  Lei Xie and
                  Yuping Wang},
  title        = {StreamVoice: Streamable Context-Aware Language Modeling for Real-time
                  Zero-Shot Voice Conversion},
  booktitle    = {{ACL} {(1)}},
  pages        = {7328--7338},
  publisher    = {Association for Computational Linguistics},
  year         = {2024}
}

@inproceedings{DBLP:conf/icassp/NingJ0WY0B24,
  author       = {Ziqian Ning and
                  Yuepeng Jiang and
                  Pengcheng Zhu and
                  Shuai Wang and
                  Jixun Yao and
                  Lei Xie and
                  Mengxiao Bi},
  title        = {Dualvc 2: Dynamic Masked Convolution for Unified Streaming and Non-Streaming
                  Voice Conversion},
  booktitle    = {{ICASSP}},
  pages        = {11106--11110},
  publisher    = {{IEEE}},
  year         = {2024}
}

@inproceedings{DBLP:conf/iccv/PeeblesX23,
  author       = {William Peebles and
                  Saining Xie},
  title        = {Scalable Diffusion Models with Transformers},
  booktitle    = {{ICCV}},
  pages        = {4172--4182},
  publisher    = {{IEEE}},
  year         = {2023}
}

@inproceedings{DBLP:conf/interspeech/AISHELL-3,
  author       = {Yao Shi and
                  Hui Bu and
                  Xin Xu and
                  Shaoji Zhang and
                  Ming Li},
  title        = {{AISHELL-3:} {A} Multi-Speaker Mandarin {TTS} Corpus},
  booktitle    = {Interspeech},
  pages        = {2756--2760},
  publisher    = {{ISCA}},
  year         = {2021}
}

@inproceedings{DBLP:conf/slt/Emilia,
  author       = {Haorui He and
                  Zengqiang Shang and
                  Chaoren Wang and
                  Xuyuan Li and
                  Yicheng Gu and
                  Hua Hua and
                  Liwei Liu and
                  Chen Yang and
                  Jiaqi Li and
                  Peiyang Shi and
                  Yuancheng Wang and
                  Kai Chen and
                  Pengyuan Zhang and
                  Zhizheng Wu},
  title        = {Emilia: An Extensive, Multilingual, and Diverse Speech Dataset For
                  Large-Scale Speech Generation},
  booktitle    = {{SLT}},
  pages        = {885--890},
  publisher    = {{IEEE}},
  year         = {2024}
}

@inproceedings{DBLP:conf/icassp/Dnsmos,
  author       = {Chandan K. A. Reddy and
                  Vishak Gopal and
                  Ross Cutler},
  title        = {Dnsmos {P.835:} {A} Non-Intrusive Perceptual Objective Speech Quality
                  Metric to Evaluate Noise Suppressors},
  booktitle    = {{ICASSP}},
  pages        = {886--890},
  publisher    = {{IEEE}},
  year         = {2022}
}

@inproceedings{DBLP:conf/icassp/Fast-U2++,
  author       = {Chengdong Liang and
                  Xiao{-}Lei Zhang and
                  Binbin Zhang and
                  Di Wu and
                  Shengqiang Li and
                  Xingchen Song and
                  Zhendong Peng and
                  Fuping Pan},
  title        = {Fast-U2++: Fast and Accurate End-to-End Speech Recognition in Joint
                  CTC/Attention Frames},
  booktitle    = {{ICASSP}},
  pages        = {1--5},
  publisher    = {{IEEE}},
  year         = {2023}
}

@inproceedings{DBLP:conf/interspeech/WeNet,
  author       = {Zhuoyuan Yao and
                  Di Wu and
                  Xiong Wang and
                  Binbin Zhang and
                  Fan Yu and
                  Chao Yang and
                  Zhendong Peng and
                  Xiaoyu Chen and
                  Lei Xie and
                  Xin Lei},
  title        = {WeNet: Production Oriented Streaming and Non-Streaming End-to-End
                  Speech Recognition Toolkit},
  booktitle    = {Interspeech},
  pages        = {4054--4058},
  publisher    = {{ISCA}},
  year         = {2021}
}

@inproceedings{DBLP:conf/icassp/WENETSPEECH,
  author       = {Binbin Zhang and
                  Hang Lv and
                  Pengcheng Guo and
                  Qijie Shao and
                  Chao Yang and
                  Lei Xie and
                  Xin Xu and
                  Hui Bu and
                  Xiaoyu Chen and
                  Chenchen Zeng and
                  Di Wu and
                  Zhendong Peng},
  title        = {{WENETSPEECH:} {A} 10000+ Hours Multi-Domain Mandarin Corpus for Speech
                  Recognition},
  booktitle    = {{ICASSP}},
  pages        = {6182--6186},
  publisher    = {{IEEE}},
  year         = {2022}
}

@inproceedings{DBLP:conf/interspeech/ECAPA-TDNN,
  author       = {Brecht Desplanques and
                  Jenthe Thienpondt and
                  Kris Demuynck},
  title        = {{ECAPA-TDNN:} Emphasized Channel Attention, Propagation and Aggregation
                  in {TDNN} Based Speaker Verification},
  booktitle    = {{INTERSPEECH}},
  pages        = {3830--3834},
  publisher    = {{ISCA}},
  year         = {2020}
}

@inproceedings{DBLP:conf/iclr/Vocos,
  author       = {Hubert Siuzdak},
  title        = {Vocos: Closing the gap between time-domain and Fourier-based neural
                  vocoders for high-quality audio synthesis},
  booktitle    = {{ICLR}},
  publisher    = {OpenReview.net},
  year         = {2024}
}

@inproceedings{DBLP:conf/icassp/HussainNHLG23,
  author       = {Shehzeen Hussain and
                  Paarth Neekhara and
                  Jocelyn Huang and
                  Jason Li and
                  Boris Ginsburg},
  title        = {{ACE-VC:} Adaptive and Controllable Voice Conversion Using Explicitly
                  Disentangled Self-Supervised Speech Representations},
  booktitle    = {{ICASSP}},
  pages        = {1--5},
  publisher    = {{IEEE}},
  year         = {2023}
}

@inproceedings{DBLP:conf/icassp/LiLL23,
  author       = {Dayong Li and
                  Xian Li and
                  Xiaofei Li},
  title        = {{DVQVC:} An Unsupervised Zero-Shot Voice Conversion Framework},
  booktitle    = {{ICASSP}},
  pages        = {1--5},
  publisher    = {{IEEE}},
  year         = {2023}
}

@inproceedings{DBLP:conf/icassp/SEF-VC,
  author       = {Junjie Li and
                  Yiwei Guo and
                  Xie Chen and
                  Kai Yu},
  title        = {{SEF-VC:} Speaker Embedding Free Zero-Shot Voice Conversion with Cross
                  Attention},
  booktitle    = {{ICASSP}},
  pages        = {12296--12300},
  publisher    = {{IEEE}},
  year         = {2024}
}

@inproceedings{DBLP:conf/icassp/LuoD24,
  author       = {Yin{-}Jyun Luo and
                  Simon Dixon},
  title        = {Posterior Variance-Parameterised Gaussian Dropout: Improving Disentangled
                  Sequential Autoencoders for Zero-Shot Voice Conversion},
  booktitle    = {{ICASSP}},
  pages        = {11676--11680},
  publisher    = {{IEEE}},
  year         = {2024}
}

@inproceedings{DBLP:conf/icassp/KimKCNMC25,
  author       = {Jaehun Kim and
                  Ji{-}Hoon Kim and
                  Yeunju Choi and
                  Tan Dat Nguyen and
                  Seongkyu Mun and
                  Joon Son Chung},
  title        = {AdaptVC: High Quality Voice Conversion with Adaptive Learning},
  booktitle    = {{ICASSP}},
  pages        = {1--5},
  publisher    = {{IEEE}},
  year         = {2025}
}

@inproceedings{DBLP:conf/icassp/ChoiP25,
  author       = {Ha{-}Yeong Choi and
                  Jaehan Park},
  title        = {VoicePrompter: Robust Zero-Shot Voice Conversion with Voice Prompt
                  and Conditional Flow Matching},
  booktitle    = {{ICASSP}},
  pages        = {1--5},
  publisher    = {{IEEE}},
  year         = {2025}
}

@inproceedings{DBLP:conf/iclr/LipmanCBNL23,
  author       = {Yaron Lipman and
                  Ricky T. Q. Chen and
                  Heli Ben{-}Hamu and
                  Maximilian Nickel and
                  Matthew Le},
  title        = {Flow Matching for Generative Modeling},
  booktitle    = {{ICLR}},
  publisher    = {OpenReview.net},
  year         = {2023}
}

@inproceedings{DBLP:conf/iclr/HuSWALWWC22,
  author       = {Edward J. Hu and
                  Yelong Shen and
                  Phillip Wallis and
                  Zeyuan Allen{-}Zhu and
                  Yuanzhi Li and
                  Shean Wang and
                  Lu Wang and
                  Weizhu Chen},
  title        = {LoRA: Low-Rank Adaptation of Large Language Models},
  booktitle    = {{ICLR}},
  publisher    = {OpenReview.net},
  year         = {2022}
}

@inproceedings{bgmvc,
  author       = {Jixun Yao and
                  Yi Lei and
                  Qing Wang and
                  Pengcheng Guo and
                  Ziqian Ning and
                  Lei Xie and
                  Hai Li and
                  Junhui Liu and
                  Danming Xie},
  title        = {Preserving background sound in noise-robust voice conversion via multi-task
                  learning},
  booktitle    = {Proc. ICASSP},
  pages        = {1--5},
  publisher    = {{IEEE}},
  year         = {2023},
}

@inproceedings{expressive-vc,
  author       = {Ziqian Ning and
                  Qicong Xie and
                  Pengcheng Zhu and
                  Zhichao Wang and
                  Liumeng Xue and
                  Jixun Yao and
                  Lei Xie and
                  Mengxiao Bi},
  title        = {Expressive-VC: Highly Expressive Voice Conversion with Attention Fusion
                  of Bottleneck and Perturbation Features},
  booktitle    = {Proc. ICASSP},
  pages        = {1--5},
  publisher    = {{IEEE}},
  year         = {2023},
}

@article{DBLP:journals/spl/WangCWXW24,
  author       = {Zhichao Wang and
                  Yuanzhe Chen and
                  Xinsheng Wang and
                  Lei Xie and
                  Yuping Wang},
  title        = {StreamVoice+: Evolving Into End-to-End Streaming Zero-Shot Voice Conversion},
  journal      = {{IEEE} Signal Process. Lett.},
  volume       = {31},
  pages        = {3000--3004},
  year         = {2024}
}

@article{DBLP:journals/corr/seedvc,
  author       = {Songting Liu},
  title        = {Zero-shot Voice Conversion with Diffusion Transformers},
  journal      = {CoRR},
  volume       = {abs/2411.09943},
  year         = {2024}
}

@article{DBLP:journals/corr/MoonCast,
  author       = {Zeqian Ju and
                  Dongchao Yang and
                  Jianwei Yu and
                  Kai Shen and
                  Yichong Leng and
                  Zhengtao Wang and
                  Xu Tan and
                  Xinyu Zhou and
                  Tao Qin and
                  Xiangyang Li},
  title        = {MoonCast: High-Quality Zero-Shot Podcast Generation},
  journal      = {CoRR},
  volume       = {abs/2503.14345},
  year         = {2025}
}

@article{DBLP:journals/corr/MeanFlows,
  author       = {Zhengyang Geng and
                  Mingyang Deng and
                  Xingjian Bai and
                  J. Zico Kolter and
                  Kaiming He},
  title        = {Mean Flows for One-step Generative Modeling},
  journal      = {CoRR},
  volume       = {abs/2505.13447},
  year         = {2025}
}

@article{DBLP:journals/corr/HiFiSinger,
  author       = {Jiawei Chen and
                  Xu Tan and
                  Jian Luan and
                  Tao Qin and
                  Tie{-}Yan Liu},
  title        = {HiFiSinger: Towards High-Fidelity Neural Singing Voice Synthesis},
  journal      = {CoRR},
  volume       = {abs/2009.01776},
  year         = {2020}
}

@article{DBLP:journals/corr/APT,
  author       = {Shanchuan Lin and
                  Xin Xia and
                  Yuxi Ren and
                  Ceyuan Yang and
                  Xuefeng Xiao and
                  Lu Jiang},
  title        = {Diffusion Adversarial Post-Training for One-Step Video Generation},
  journal      = {CoRR},
  volume       = {abs/2501.08316},
  year         = {2025}
}

@article{DBLP:journals/corr/Seed-tts,
  author       = {Philip Anastassiou and
                  Jiawei Chen and
                  Jitong Chen and
                  Yuanzhe Chen and
                  Zhuo Chen and
                  Ziyi Chen and
                  Jian Cong and others},
  title        = {Seed-TTS: {A} Family of High-Quality Versatile Speech Generation Models},
  journal      = {CoRR},
  volume       = {abs/2406.02430},
  year         = {2024}
}

@article{DBLP:journals/taslp/ZhangLD20,
  author       = {Jing{-}Xuan Zhang and
                  Zhen{-}Hua Ling and
                  Li{-}Rong Dai},
  title        = {Non-Parallel Sequence-to-Sequence Voice Conversion With Disentangled
                  Linguistic and Speaker Representations},
  journal      = {{IEEE} {ACM} Trans. Audio Speech Lang. Process.},
  volume       = {28},
  pages        = {540--552},
  year         = {2020}
}

@article{voiceprivacy2022,
  author       = {Jixun Yao and
                  Qing Wang and
                  Li Zhang and
                  Pengcheng Guo and
                  Yuhao Liang and
                  Lei Xie},
  title        = {{NWPU-ASLP} System for the VoicePrivacy 2022 Challenge},
  journal      = {CoRR},
  volume       = {abs/2209.11969},
  year         = {2022},
  eprinttype    = {arXiv},
  eprint       = {2209.11969},
}

\end{document}